\documentstyle[12pt,epsfig,graphicx]{article} 
\makeatletter \def\@cite#1#2{{[{#1}]\if@tempswa\typeout {IJCGA
warning: optional citation argument ignored: `#2'} \fi}}

\newcount\@tempcntc
\def\@citex[#1]#2{\if@filesw\immediate\write\@auxout{\string\citation{#2}}\fi
  \@tempcnta\z@\@tempcntb\m@ne\def\@citea{}\@cite{\@for\@citeb:=#2\do
    {\@ifundefined
       {b@\@citeb}{\@citeo\@tempcntb\m@ne\@citea\def\@citea{,}{\bf ?}\@warning
       {Citation `\@citeb' on page \thepage \space undefined}}%
    {\setbox\z@\hbox{\global\@tempcntc0\csname b@\@citeb\endcsname\relax}%
     \ifnum\@tempcntc=\z@ \@citeo\@tempcntb\m@ne
       \@citea\def\@citea{,}\hbox{\csname b@\@citeb\endcsname}%
     \else
     \advance\@tempcntb\@ne
      \ifnum\@tempcntb=\@tempcntc
      \else\advance\@tempcntb\m@ne\@citeo
      \@tempcnta\@tempcntc\@tempcntb\@tempcntc\fi\fi}}\@citeo}{#1}}
\def\@citeo{\ifnum\@tempcnta>\@tempcntb\else\@citea\def\@citea{,}%
  \ifnum\@tempcnta=\@tempcntb\the\@tempcnta\else
   {\advance\@tempcnta\@ne\ifnum\@tempcnta=\@tempcntb \else 
\def\@citea{--}\fi
    \advance\@tempcnta\m@ne\the\@tempcnta\@citea\the\@tempcntb}\fi\fi}
\makeatother

\def\boxit#1{\leavevmode\thinspace\hbox{\vrule\vtop{\vbox{\hrule%
        \vskip3pt\kern1pt\hbox{\vphantom{\bf/}\thinspace\thinspace%
        {\bf#1}\thinspace\thinspace}}\kern1pt\vskip3pt\hrule}\vrule}%
        \thinspace}
\def\Boxit#1{\noindent\vbox{\hrule\hbox{\vrule\kern3pt\vbox{
\advance\hsize-7pt\vskip-\parskip\kern3pt\bf#1 \hbox{\vrule height0pt
depth\dp\strutbox width0pt} \kern3pt}\kern3pt\vrule}\hrule}}

\newcommand{\gsim}{\lower.7ex\hbox{$\;\stackrel{\textstyle>}{\sim}\;$}}
\newcommand{\lsim}{\lower.7ex\hbox{$\;\stackrel{\textstyle<}{\sim}\;$}}
\newcommand{\be}{\begin{equation}} \newcommand{\ee}{\end{equation}}
\newcommand{\beq}{\begin{equation}} \newcommand{\eeq}{\end{equation}}
\newcommand{\bea}{\begin{eqnarray}} \newcommand{\eea}{\end{eqnarray}}

\def\baselinestretch{1}
\begin{document}
\catcode`@=11 \newtoks\@stequation
\def\subequations{\refstepcounter{equation}%
\edef\@savedequation{\the\c@equation}%
\@stequation=\expandafter{\theequation}
\edef\@savedtheequation{\the\@stequation}
\edef\oldtheequation{\theequation}
\def\theequation{\oldtheequation\alph{equation}}}
\def\endsubequations{\setcounter{equation}{\@savedequation}%
\@stequation=\expandafter{\@savedtheequation}%
\edef\theequation{\the\@stequation}\global\@ignoretrue

\noindent} \catcode`@=12
\begin{titlepage}

\title{{\bf  Antimatter Signatures \\of Gravitino Dark Matter Decay}}
\vskip3in \author{{\bf Alejandro Ibarra} and
{\bf David Tran\footnote{\baselineskip=16pt {\small E-mail addresses: {\tt
alejandro.ibarra@desy.de, david.tran@desy.de}}}}
\hspace{3cm}\\ \vspace{0.1cm}
{\small DESY,  Theory Group, Notkestrasse 85, D-22603 Hamburg, Germany}.
}  \date{}  \maketitle  \def\baselinestretch{1.15}
\begin{abstract}
\noindent 
The scenario of gravitino dark matter with broken $R$-parity
naturally reconciles three paradigms that, albeit very well
motivated separately, seem to be in mutual conflict: supersymmetric
dark matter, thermal leptogenesis and standard Big Bang nucleosynthesis. 
Interestingly enough, the products of the gravitino decay
could be observed, opening the possibility of indirect
detection of gravitino dark matter. 
In this paper, we compute the positron and the antiproton fluxes
from gravitino decay. We find that a gravitino with a mass of
$m_{3/2}\sim 150$ GeV and a lifetime of $\tau_{3/2}\sim 10^{26}$ s
could simultaneously explain the 
EGRET anomaly in the extragalactic diffuse gamma ray background
and the HEAT excess in the positron fraction. However,
the predicted antiproton flux tends to be too large, although 
the prediction suffers from large uncertainties and might be compatible 
with present observations for certain choices of propagation parameters.
\end{abstract}

\thispagestyle{empty}
\vspace*{0.2cm} \leftline{April 2008} \leftline{}

\vskip-15cm \rightline{DESY 08-048} 
\vskip3in

\end{titlepage}
\setcounter{footnote}{0} \setcounter{page}{1}
\newpage
\baselineskip=20pt

\noindent

\section{Introduction}

Models with local supersymmetry predict the existence of a
particle with extremely weak interactions: the gravitino.
In contrast to the supersymmetric partners of the Standard Model
particles, whose masses are expected to lie in the electroweak
domain, the gravitino can have a mass ranging between 
a few eV and several TeV without conflicting with any laboratory
experiment. Therefore, the gravitino can very naturally be
the lightest supersymmetric particle (LSP), and if it is
sufficiently long-lived, it could constitute the dark matter
of the Universe~\cite{Pagels:1981ke}.

Gravitinos were produced in the early Universe by
scatterings in the thermal plasma, but did not subsequently
annihilate due to their extremely weak interactions.
Therefore, a relic population of gravitinos is
expected in the present Universe with a density
given by~\cite{Bolz:2000fu}
\begin{equation}
\Omega_{3/2} h^2\simeq 0.27
    \left(\frac{T_R}{10^{10}\,{\rm GeV}}\right)
    \left(\frac{100 \,{\rm GeV}}{m_{3/2}}\right)
    \left(\frac{m_{\widetilde g}}{1\,{\rm TeV}}\right)^2\;,
\label{relic-abundance}
\end{equation}
where $T_R$ is the reheating temperature of the Universe,
$m_{3/2}$ is the gravitino mass and $m_{\widetilde g}$ is the 
gluino mass. In predicting the relic abundance of gravitinos,
the main uncertainty arises from our ignorance of the thermal 
history of the Universe before Big Bang nucleosynthesis (BBN) and
in particular of the reheating temperature after inflation.
However, we have strong indications that the Universe was very hot 
after inflation. Namely, the discovery of neutrino masses about ten
years ago provided strong support to leptogenesis
as the explanation for the observed baryon asymmetry of
the Universe. This mechanism can reproduce the observed
baryon asymmetry very naturally if the reheating
temperature of the Universe was above $10^9$ GeV~\cite{Davidson:2002qv}.
Therefore, following Eq.~(\ref{relic-abundance}), the gravitino could
constitute the dark matter if $m_{3/2}\gsim 10$ GeV
for a gluino mass $m_{\widetilde g}\simeq 1$ TeV, which 
is consistent with the assumption that the gravitino is the lightest
supersymmetric particle.

Remarkably, the conjectures of a reheating temperature of the Universe larger
than $10^9\,{\rm GeV}$ and a gravitino mass larger than a few GeV
can naturally solve two of the most long-standing problems
in cosmology: the nature of the dark matter and
the origin of the baryon asymmetry of the Universe. Nevertheless,
this picture is not exempt from problems. If $R$-parity is
exactly conserved, the next-to-LSP (NSLP) can only decay gravitationally
into gravitinos and Standard Model particles with a lifetime
\begin{equation}
\tau_{\rm NLSP}\simeq 9~{\rm days}
\left(\frac{m_{3/2}}{10~{\rm GeV}}\right)^2
\left(\frac{150{\rm GeV}}{m_{\rm NLSP}}\right)^5\;.
\end{equation}
Then, the NLSP is typically present during and after Big Bang
nucleosynthesis, jeopardizing the successful predictions
of the standard nucleosynthesis scenario. This is in fact
the case for the most likely candidates for the NLSP: the lightest
neutralino and the right-handed stau (or more generically, any
negatively charged particle, such as the chargino). More precisely, 
when the NLSP is the neutralino, the hadrons produced in the
neutralino decays typically dissociate the primordial 
elements~\cite{Kawasaki:2004qu},
yielding abundances in conflict with observations.
On the other hand, when the NLSP is a charged particle, $X^-$, the formation 
of the bound state $(^4{\rm He}\,X^-)$ catalyzes the production 
of $^6$Li~\cite{Pospelov:2006sc} leading to an overproduction 
of $^6$Li by a factor $300 - 600$~\cite{Hamaguchi:2007mp}.

Although the scenario depicted above is the most widely studied,
it is not the only possibility. Indeed, several alternatives
have been proposed that yield a thermal history of the 
Universe consistent with the observed 
relic density of dark matter, successful leptogenesis and
successful Big Bang nucleosynthesis. For instance, in some
specific supersymmetric models the NLSP can be a 
sneutrino~\cite{Kanzaki:2006hm} or a stop~\cite{DiazCruz:2007fc},
whose late decays do not substantially affect 
the predictions of Big Bang  nucleosynthesis.
Another possibility is to assume some amount
of entropy production after NLSP decoupling, which dilutes the
NLSP abundance~\cite{Pradler:2006hh}. 
Finally, if $R$-parity is not exactly
conserved, the NLSP can decay into two Standard model particles
well before the onset of Big Bang nucleosynthesis, avoiding
the BBN constraints altogether~\cite{Buchmuller:2007ui}. 
This is the scenario that we will adopt in this paper.

When $R$-parity is not imposed, the superpotential of the
Minimal Supersymmetric Standard Model (MSSM) reads~\cite{Barbier:2004ez}
\begin{eqnarray}
\label{Rp-superpotential}
W=W_{R_p} + \frac{1}{2} \lambda_{ijk} L_i L_j e^c_k
+ \lambda^{\prime}_{ijk}\, L_i Q_j d^c_k 
+ \frac{1}{2} \lambda^{\prime\prime}_{ijk} u^c_i d^c_j d^c_k + 
\mu_i L_i H_u\;,
\end{eqnarray}
where $W_{R_p}$ is the familiar superpotential with conserved
$R$-parity. Present laboratory experiments very severely restrict the
size of the $R$-parity breaking couplings. For instance, when
the soft masses are $\sim 100$ GeV,
proton stability requires $\lambda^\prime_{11k}\lambda^{\prime\prime}_{11k}
\lsim 10^{-27}$, and 
the non-observation of the lepton flavor violating
process $\mu {\rm Ti}\rightarrow e {\rm Ti}$ requires 
$\lambda_{1k2} \lambda^{\prime}_{k11}\lsim 4\times 10^{-8}$, for
$k=1,2,3$. An exhaustive list of the laboratory constraints on 
the $R$-parity violating couplings can be found in~\cite{Allanach:1999ic}.

In addition to the laboratory
upper bounds, there also exists an allowed {\it window} for 
the $R$-parity violating Yukawa couplings stemming from cosmology. 
If the $R$-parity violating interactions had been
in thermal equilibrium before the electroweak phase transition,
any preexisting baryon or lepton asymmetry would have been erased. 
Therefore, successful leptogenesis can only be achieved if the 
out-of-equilibrium condition
$\lambda, \lambda^{\prime}, \lambda^{\prime\prime}\lsim 10^{-7}$
is satisfied~\cite{Campbell:1992jd}. These bounds are 
sufficient but not necessary conditions
and could be relaxed for some specific flavor structures.
On the other hand, successful Big Bang nucleosynthesis
is guaranteed if the NLSP lifetime is shorter than $\sim 10^3\;{\rm s}$,
which yields a lower bound on the $R$-parity breaking Yukawa couplings. 
For instance, when the NLSP is a right-handed stau, it can
decay via $\widetilde \tau_R\rightarrow \mu \nu_\tau$ through the coupling
$\lambda_{323}$ with lifetime
\begin{equation}
\tau_{\widetilde \tau} \simeq 10^3\,{\rm s} 
\left(\frac{\lambda_{323}}{10^{-14}}\right)^{-2}
\left(\frac{m_{\widetilde \tau}}{{100~{\rm GeV}}}\right)^{-1}.
\end{equation}
Therefore, even a tiny amount of $R$-parity violation,  
$\lambda_{323}\gsim 10^{-14}$, is enough to deplete the population
of stau NLSPs at the time of BBN down to harmless 
levels~\cite{Buchmuller:2007ui}. A similar argument applies 
for the case of a neutralino NLSP with analogous conclusions.

When $R$-parity is not exactly conserved, the gravitino LSP
is no longer stable. Nevertheless, the gravitino decay rate
is doubly suppressed by the Planck mass and by the
small $R$-parity violation, yielding~\cite{Takayama:2000uz,Bertone:2007aw}
\begin{equation}
\label{grav-lifetime}
\tau_{3/2}\ \simeq \ 10^{23}\, {\rm s}\  
\left(\frac{\lambda}{10^{-7}}\right)^{-2}
\left(\frac{m_{3/2}}{100~{\rm GeV}}\right)^{-3}\;.
\end{equation}
Therefore, for the range of $R$-parity violating couplings
favored by cosmology,  $10^{-14}\lsim$ $\lambda, \lambda' \lsim 10^{-7}$,
the gravitino lifetime ranges between $10^{23}$ and $10^{37}\;{\rm s}$
for $m_{3/2}=100\;{\rm GeV}$, which exceeds the age of the Universe 
by many orders of magnitude. Hence, even though the gravitino
is not absolutely stable, it is stable enough to constitute 
a viable candidate for the dark matter of the Universe, 
while preserving the attractive
features of  the standard Big Bang nucleosynthesis scenario 
and thermal leptogenesis.

Interestingly, the gravitino decay products could be observed
as a contribution to the flux of cosmic gamma rays, 
positrons, antiprotons and neutrinos, opening the possibility 
of indirect detection of gravitino dark matter. 
We computed in \cite{Ibarra:2007wg} the gamma ray
spectrum from gravitino decay, and we found that the 
anomaly in the extragalactic gamma ray flux reported by Strong {\it et al.} 
between 2-10 GeV~\cite{smr05} in the EGRET observations~\cite{egret}
could be qualitatively explained by the 
decay of gravitino dark matter with a mass of
$m_{3/2}\simeq 150\;{\rm GeV}$ and a lifetime of 
$\tau_{3/2}\simeq 1.3\times 10^{26}\;{\rm s}$\footnote{The lifetime 
quoted in \cite{Ibarra:2007wg},
$\tau_{3/2}\simeq 1.6\times 10^{26}\;{\rm s}$, corresponds to
a local halo density $\rho_\odot=0.38\;{\rm GeV}/{\rm cm}^3$.
Here we have adopted the more standard value 
$\rho_\odot=0.30\,{\rm GeV}/{\rm cm}^3$~\cite{Bergstrom:1997fj}. 
Note that the flux 
of particles from gravitino decay is proportional to 
$\rho_\odot/\tau_{3/2}$; therefore the uncertainty in the 
value of the local halo density translates into an uncertainty 
in the gravitino lifetime.}. 
The expected anisotropy in the diffuse gamma ray flux was also 
found to be consistent with the EGRET observations.

Motivated by this result, in this paper we compute
the predicted fluxes of positrons and antiprotons from gravitino decay
for the same set of parameters, as independent tests of this 
scenario\footnote{The predictions for the neutrino flux will 
be presented elsewhere~\cite{CGIT}.}. The flux of positrons 
has been measured by a series of experiments:
HEAT~\cite{Barwick:1997ig}, CAPRICE~\cite{CAPRICE}, 
MASS~\cite{Grimani:2002yz} and AMS-01~\cite{Aguilar:2007yf}. 
Clearly, if gravitino decay is the explanation for the extragalactic
EGRET anomaly, our predicted positron flux should not exceed the 
measured one. Although the measurements still suffer from large
uncertainties, it is intriguing that they seem to point 
to an excess of positrons at energies larger than 
7 GeV, which is precisely the energy range where we expect
a contribution of positrons from gravitino decay.
On the other hand, the antiproton
flux has been measured by BESS~\cite{Orito:1999re}, 
IMAX~\cite{Mitchell:1996bi} and WiZard/CAPRICE~\cite{Boezio:1997ec}. 
The measurements do not show
any deviation from the predictions by conventional astrophysical 
models of spallation of cosmic rays on the Milky Way disk.
Therefore, the viability of our scenario requires that the total
antiproton flux lie below the astrophysical background.
Future antimatter experiments such as PAMELA~\cite{Picozza:2006nm}
or AMS-02~\cite{Barao:2004ik} will provide  very precise
measurements of the spectra of positrons and antiprotons
and will provide important constraints on the scenario 
of decaying dark matter.

In Section 2 we derive the source term for positrons and antiprotons
from the decay of gravitinos in the Milky Way halo. Both species, 
being electrically charged, propagate through the halo in a 
complicated way that we simulate by means of a conventional 
diffusion model. In section 3 we discuss and solve the diffusion
equation for positrons and antiprotons. Finally, we present
our conclusions and an outlook in section 4.

\section{Source Term}

We will assume that the Milky Way dark matter halo is 
populated by gravitinos with mass $m_{3/2}$, their distribution following a 
density profile $\rho(\vec{r})$, where $\vec{r}$ denotes
the position with respect to the center of the Galaxy.
The dark matter distribution is usually parametrized as a spherically
symmetric profile
\begin{equation}
\rho(r)=\frac{\rho_0}{(r/r_c)^\gamma 
[1+(r/r_c)^\alpha]^{(\beta-\gamma)/\alpha}}\;,
\end{equation}
where $r=|\vec{r}|$ and the parameters $\alpha$, $\beta$, $\gamma$ and
$r_c$ are listed in Table~\ref{tab:halomodels} for some commonly 
used halo models. Finally, $\rho_0$ is a parameter that is adjusted 
to yield a local halo density of
$\rho(r_\odot)=0.30\,{\rm GeV}/{\rm cm}^3$~\cite{Bergstrom:1997fj},
with $r_\odot = 8.5 ~\rm{kpc}$ being the distance of the Sun to the 
Galactic center.

\begin{table}[t]
\begin{center}
\begin{tabular}{|c|cccc|}
\hline
Halo model & $\alpha$ & $\beta$ & $\gamma$ & $r_c$ (kpc)\\
\hline
Navarro, Frenk, White~\cite{Navarro:1995iw}& 1 & 3 & 1 & 20\\
Isothermal & 2 & 2 & 0 & 3.5 \\
Moore~\cite{Moore:1999gc}& 1.5 & 3 & 1.5 &28\\
\hline
\end{tabular}
\caption{\label{tab:halomodels} \small 
Parameters characterizing some commonly used halo models.}
\end{center}
\end{table}

Gravitinos at $\vec{r}$ eventually decay with lifetime $\tau_{3/2}$
producing antimatter at a rate per unit energy and unit volume 
given by
\begin{equation}
Q(E,\vec{r})=\frac{\rho(\vec{r})}{m_{3/2}\tau_{3/2}}\frac{dN}{dE}\;,
\label{source-term}
\end{equation}
where $dN/dE$ is the energy spectrum of antiparticles
produced in the decay.

If the gravitino is lighter than the $W^\pm$ gauge bosons,
the main decay channel is $\psi_{3/2}\rightarrow \nu\gamma$,
which does not produce antimatter. On the other hand, if
it is heavier than the gauge bosons, the decay channels
$\psi_{3/2}\rightarrow W^\pm \ell^\mp$ and
$\psi_{3/2}\rightarrow Z^0 \nu$ are kinematically accessible
and produce antimatter\footnote{The size of the 
$R$-parity violating couplings for the third generation is
expected to be larger than for the first and second 
generations. Consequently, the charged lepton and
the neutrino produced in these decays are expected
to have predominantly tau flavor.}. 
Namely, the process $\psi_{3/2}\rightarrow W^- e^+$
produces a high-energy monoenergetic positron. On the other hand,
the antimuon and antitau produced in the processes
$\psi_{3/2}\rightarrow W^- \mu^+, W^- \tau^+$ generate a continuous
spectrum of positrons in their decays. Lastly, the main contribution
comes from the fragmentation
of the $W^\pm$ and the $Z^0$ gauge bosons, which produce 
a continuous spectrum of positrons (mainly from
$\pi^+$ decay) and antiprotons that we have obtained using the 
event generator PYTHIA 6.4~\cite{Sjostrand:2006za}. Thus, the total energy
spectrum of antiparticles reads:
\begin{eqnarray}
\frac{dN}{dE}\simeq
{\rm BR}(\psi_{3/2}\rightarrow W \ell)\frac{dN^{W\ell}}{dE}+
{\rm BR}(\psi_{3/2}\rightarrow Z^0 \nu)\frac{dN^{Z\nu}}{dE} \,. 
\label{inj-spectrum}
\end{eqnarray}
where $dN^{W\ell}/dE$ and $dN^{Z\nu}/dE$ denote the energy spectra
of antiparticles produced in the indicated decay channel, 
which depend only on the gravitino mass through the total
available energy. 

On the other hand, the branching ratios of the relevant decay 
channels can be straightforwardly computed from the expressions
for the gravitino decay rates in~\cite{Ibarra:2007wg}. 
The result is:
\begin{eqnarray}
{\rm BR}(\psi_{3/2}\rightarrow W \ell) &=& 
\frac{2|U_{\tilde W \ell}|^2\,f(\frac{M_W}{m_{3/2}})}
{|U_{\tilde \gamma\nu}|^2+
2|U_{\tilde W \ell}|^2\,f(\frac{M_W}{m_{3/2}})+
|U_{\tilde Z\nu}|^2\,f(\frac{M_Z}{m_{3/2}})}\;,
\nonumber \\
{\rm BR}(\psi_{3/2}\rightarrow Z^0 \nu)&=& 
\frac{|U_{\tilde Z\nu}|^2\,f(\frac{M_Z}{m_{3/2}})}
{|U_{\tilde \gamma\nu}|^2+
2|U_{\tilde W \ell}|^2\,f(\frac{M_W}{m_{3/2}})+
|U_{\tilde Z\nu}|^2\,f(\frac{M_Z}{m_{3/2}})}\;,
\end{eqnarray}
where $f(x)=1-\frac{4}{3}x^2+\frac{1}{3}x^8$
and $U_{\tilde \gamma \nu}$, $U_{\tilde W \ell}$, $U_{\tilde Z\nu}$
denote the mixings photino-neutrino, 
charged wino-charged lepton and  zino-neutrino, respectively, which satisfy 
the following relations:
\begin{eqnarray}
|U_{\widetilde \gamma \nu}|&\simeq& 
\left[\frac{(M_2-M_1) \sin\theta_W \cos\theta_W }
{ M_1 \cos^2\theta_W + M_2  \sin^2\theta_W}\right]
|U_{\widetilde Z \nu}|\;, \nonumber \\
|U_{\widetilde W \ell}|&\simeq&
\sqrt{2}\cos\theta_W\frac{M_1 \sin^2\theta_W + M_2 \cos^2\theta_W}
{M_2}|U_{\widetilde Z \nu}| \;.
\label{UWtau2}
\end{eqnarray}
In this expression, $M_1$ and $M_2$ are the $U(1)_Y$ and 
$SU(2)_L$ gaugino masses
at low energies and $\theta_W$ is the weak mixing angle. 
It is commonly assumed that the gaugino masses unify at the
Grand Unified Scale, $M_X=2\times 10^{16}\,{\rm GeV}$. Under
this assumption, the ratio between the  gaugino masses
at low energies is predicted to be $M_2/M_1\simeq 1.9$, which yields
\begin{equation}
|U_{\widetilde \gamma \nu}|:|U_{\widetilde Z \nu}|:|U_{\widetilde W \ell}|
\simeq 1:3.2:3.5\;.
\end{equation}
Therefore, the gravitino branching ratios in the different decay modes
depend only on the gravitino mass (see Table~\ref{tab:BRs}).

We conclude that, under the assumption of gaugino mass
universality, the total energy spectrum of antiparticles from
gravitino decay, $dN/dE$,  depends {\it exclusively}
on the gravitino mass. This makes our scenario very predictive:
for a given halo model the source term $Q(E,\vec{r})$ 
depends on only two unknown parameters, namely the gravitino mass  
and the gravitino lifetime; the former determines the spectral shape
of the source function and the latter the normalization.

\begin{table}[t]
\begin{center}
\begin{tabular}{|c|ccc|}
\hline
$m_{3/2}$ & ${\rm BR}(\psi_{3/2}\rightarrow \gamma \nu)$
&${\rm BR}(\psi_{3/2}\rightarrow W \ell)$ &
${\rm BR}(\psi_{3/2}\rightarrow Z^0 \nu)$\\
\hline
10 GeV & 1    & 0    & 0    \\
85 GeV & 0.66 & 0.34 & 0    \\
100 GeV& 0.16 & 0.76 & 0.08 \\
150 GeV& 0.05 & 0.71 & 0.24 \\
250 GeV& 0.03 & 0.69 & 0.28 \\
\hline
\end{tabular}
\caption{\label{tab:BRs} \small 
Branching ratios for gravitino decay in different $R$-parity 
violating channels for different gravitino masses.}
\end{center}
\end{table}

\section{Antimatter Propagation in the Galaxy}

Antimatter propagation in the Milky Way is commonly described by
a stationary two-zone diffusion model with cylindrical boundary 
conditions~\cite{ACR}. Under this approximation, 
the number density of antiparticles
per unit kinetic energy, $f(T,\vec{r},t)$, satisfies the following
transport equation, which applies both for positrons and antiprotons:
\begin{equation}
0=\frac{\partial f}{\partial t}=
\nabla \cdot [K(T,\vec{r})\nabla f] +
\frac{\partial}{\partial T} [b(T,\vec{r}) f]
-\nabla \cdot [\vec{V_c}(\vec{r})  f]
-2 h \delta(z) \Gamma_{\rm ann} f+Q(T,\vec{r})\;.
\label{transport}
\end{equation}
The boundary conditions require the solution 
$f(T,\vec{r},t)$ to vanish at the boundary
of the diffusion zone, which is approximated by a cylinder with 
half-height $L = 1-15~\rm{kpc}$ and radius $ R = 20 ~\rm{kpc}$.

The first term on the right-hand side of the transport equation
is the diffusion
term, which accounts for the propagation through the
tangled Galactic magnetic field.
The diffusion coefficient $K(T,\vec{r})$ is assumed to be constant
throughout the diffusion zone and is parametrized by:
\begin{equation}
K(T)=K_0 \;\beta\; {\cal R}^\delta
\end{equation}
where
$\beta=v/c$ and ${\cal R}$ is the rigidity of the particle, which
is defined as the momentum in GeV per unit charge, 
${\cal R}\equiv p({\rm GeV})/Z$.
The normalization $K_0$ and the spectral index $\delta$
of the diffusion coefficient are related to the properties 
of the interstellar medium and can be determined from the 
flux measurements of other cosmic ray species, mainly from 
the Boron to Carbon (B/C) ratio~\cite{Maurin:2001sj}. 
The second term accounts for energy losses due to 
inverse Compton scattering on starlight or the cosmic microwave 
background, synchrotron radiation and ionization. 
The third term is the convection term, which accounts for
the drift of charged particles away from the 
disk induced by the Milky Way's Galactic wind. 
It has axial direction and is also assumed to be constant
inside the diffusion region: 
$\vec{V}_c(\vec{r})=V_c\; {\rm sign}(z)\; \vec{k}$.
The fourth term accounts for antimatter annihilation with rate
$\Gamma_{\rm ann}$, when it interacts with ordinary matter
in the Galactic disk,
which is assumed to be an infinitely thin disk with half-width
$h=100$ pc. 
Lastly, $Q(T,\vec{r})$ is the source term of positrons or
antiprotons which was derived in Section 2. In
this equation, reacceleration effects and non-annihilating
interactions of antimatter in the Galactic disk have been 
neglected.

The solution of the transport equation at the Solar System, 
$r=r_\odot$, $z=0$, can be formally expressed by the convolution
\begin{equation}
f(T)=\frac{1}{m_{3/2} \tau_{3/2}}
\int_0^{T{\rm max}}dT^\prime G(T,T^\prime) 
 \frac{dN(T^\prime)}{dT^\prime}\;,
\label{solution}
\end{equation}
where $T_{\rm max}=m_{3/2}$ for the case of the positrons 
and $T_{\rm max}=m_{3/2}-m_p$ for the antiprotons.
The solution is thus factorized into two parts.
The first part, given by the Green's function $G(T,T^\prime)$,
encodes all of the information about the astrophysics 
(such as the details of the halo profile and the 
complicated propagation of antiparticles in the Galaxy) 
and is universal for any decaying dark matter candidate. The
remaining part depends exclusively on the nature and properties
of the decaying dark matter candidate, namely the mass, the lifetime 
and the energy spectrum of antiparticles produced in the decay.

Finally, the flux of primary antiparticles at the Solar System
from dark matter decay is given by:
\begin{equation}
\Phi^{\rm{prim}}(T) = \frac{v}{4 \pi} f(T),
\label{flux}
\end{equation}
where $v$ is the velocity of the antimatter particle.

In the scenario we are considering the gravitino mass and 
lifetime are constrained by requiring a qualitatively good
agreement of the predicted extragalactic gamma ray flux with
the EGRET data: $m_{3/2}=150\,{\rm GeV}$ and 
$\tau_{3/2}=1.3\times 10^{26}\,{\rm s}$~\cite{Ibarra:2007wg}.
On the other hand,  the energy spectrum of antiparticles, $dN/dT$,
is determined by the well-understood physics of fragmentation.
Therefore, the only uncertainties in the computation of the
antimatter fluxes stem from the determination of the 
Green's function, {\it i.e.} from the uncertainties in the propagation
parameters and the halo profile. As we will see, the uncertainties
in the precise shape of the halo profile are not crucial for
the determination of the primary antimatter fluxes, since
the Earth receives only antimatter created within a few kpc 
from the Sun, where the different halo profiles are very similar.
On the other hand, the uncertainties in the propagation parameters 
can substantially change the predictions for the antimatter fluxes, 
even by two orders of magnitude for the antiproton flux. 

The reason for this large uncertainty is a correlation among
the diffusion parameters and the size of the diffusion zone.
Secondary cosmic rays are produced by spallation of primary cosmic rays
in the Galactic disk. Therefore, the measurement of primary and
secondary cosmic ray fluxes (particularly the Boron to Carbon ratio) 
provides information about the diffusive properties of the interstellar 
medium. Unfortunately, there exist degeneracies in the 
determination of the diffusion parameters. For instance,
an increase in the size of the diffusion zone, which allows for
a longer propagation time of cosmic rays inside the diffusion
zone before escaping, can be compensated by a simultaneous 
increase of the diffusion coefficient, which facilitates a
faster diffusion of cosmic rays away from the Galactic disk. 
However, this degeneracy does not hold for the antimatter 
fluxes from decaying dark matter, since antimatter 
is not only produced in the Galactic disk, but in the whole 
dark matter halo. Therefore, an increase in the size of the diffusion
zone translates into an increase in the number of injected primary
antiparticles, which is not compensated by the simultaneous increase of 
the diffusion coefficient. As a result, the antimatter fluxes from 
decaying dark matter can vary substantially for the range of 
astrophysical parameters which successfully reproduce the
secondary cosmic ray fluxes. The ranges of the astrophysical
parameters that are consistent with the B/C ratio and that
produce the maximal, median and minimal positron and antiproton
fluxes are listed in Tables \ref{tab:param-positron} and
\ref{tab:param-antiproton}~\cite{Delahaye:2007fr}.

Positrons and antiprotons have different properties and their
respective transport equations can be approximated by 
different limits of Eq.~(\ref{transport}), thus allowing 
simple analytic solutions. Let us discuss each case separately.

\subsection{Positron Flux}

\begin{table}[t]
\begin{center}
\begin{tabular}{|c|ccc|}
\hline
Model & $\delta$ & $K_0\,({\rm kpc}^2/{\rm Myr})$ & $L\,({\rm kpc})$ \\
\hline 
M2 & 0.55 & 0.00595 & 1  \\
MED & 0.70 & 0.0112 & 4 \\
M1 & 0.46 & 0.0765 & 15  \\
\hline
\end{tabular}
\caption{\label{tab:param-positron}\small 
Astrophysical parameters compatible with the B/C ratio that yield 
the minimum (M2), median (MED) and maximal (M1) flux of positrons.}
\end{center}
\end{table}

For the case of the positrons, Galactic convection and annihilations
in the disk can be neglected in the transport equation, which is
then simplified to:
\begin{equation}
\nabla \cdot [K(T,\vec{r})\nabla f_{e^+}] +
\frac{\partial}{\partial T} [b(T,\vec{r}) f_{e^+}]+Q(T,\vec{r})=0\;,
\label{transport-positron}
\end{equation}
where the rate of energy loss, $b(T,\vec{r})$, is assumed to be 
a spatially constant function parametrized by $b(T)=\frac{T^2}{T_0\tau_E}$, 
with $T_0=1\;{\rm GeV}$ and $\tau_E=10^{16}\;{\rm s}$.

The solution to this equation is formally given by the convolution
Eq.~(\ref{solution}). The explicit form of the Green's function 
is~\cite{Hisano:2005ec}
\begin{equation}
G_{e^+}(T,T^\prime)=\sum_{n,m=1}^\infty 
B_{nm}(T,T^\prime) 
J_0\left(\zeta_n \frac{r_\odot}{R}\right) 
\sin\left(\frac{m \pi}{2 }\right),
\label{greens-function}
\end{equation}
where $J_0$ is the zeroth-order Bessel function of the first kind, whose
successive zeros are denoted by $\zeta_n$. On the other hand,
\begin{equation}
B_{nm}(T,T^\prime)=\frac{\tau_E T_0}{T^2}
C_{nm} 
\exp\left\{\left(\frac{\zeta_n^2}{R^2} + \frac{m^2 \pi^2}{4 L^2}\right) 
\frac{K_0 \tau_E}{\delta - 1} 
\left[\left(\frac{T}{T_0}\right)^{\delta-1}
-\left(\frac{{T^\prime}}{T_0}\right)^{\delta-1}\right]\right\},
\end{equation}
with
\begin{equation}
C_{nm}=\frac{2}{J_1^2(\zeta_n)R^2 L} \int_0^R r^\prime dr^\prime 
\int_{-L}^L dz^\prime  \rho(\vec{r}\,^\prime)
J_0\left(\zeta_n \frac{r^\prime}{R}\right) 
\sin\left[\frac{m \pi}{2 L}(L-z^\prime )\right]\;,
\end{equation}
where $J_1$ is the first-order Bessel function.

The Green's function can be well approximated by the following
interpolating function, which is valid for any decaying dark 
matter particle:
\begin{equation}
G_{e^+}(T,T^\prime)\simeq\frac{10^{16}}{T^2}
e^{a+b(T^{\delta-1}-T^{\prime \delta-1})}
\theta(T^\prime-T)\,{\rm cm}^{-3}\,{\rm s}\;,
\label{interp-pos}
\end{equation}
where $T$ and $T^\prime$ are expressed in units of GeV.
The coefficients $a$ and $b$ can be found in
Table~\ref{tab:fit-positron} for the NFW profile and 
the different diffusion models listed in Table~\ref{tab:param-positron}.
This approximation works better than a 15-20\% over the whole range
of energies.
We find numerically that the Green's function is not very sensitive
to the choice of the halo profile; therefore the corresponding 
coefficients can be well approximated by Table~\ref{tab:fit-positron}.

\begin{table}[t]
\begin{center}
\begin{tabular}{|c|cc|}
 \hline
model & $a$ & $b$ \\ 
\hline
M2 & $-0.9716$ & $-10.012$  \\
MED & $-1.0203$ & $-1.4493$  \\
M1 & $-0.9809$ & $-1.1456$  \\
 \hline
\end{tabular}
\caption{\label{tab:fit-positron}\small 
Coefficients of the interpolating function Eq.~(\ref{interp-pos}) 
for the positron Green's function, assuming a NFW halo profile
and for the different diffusion models in Table~\ref{tab:param-positron}.}
\end{center}
\end{table}

\begin{figure}[t]
\begin{center}
\begin{tabular}{c}
\psfig{figure=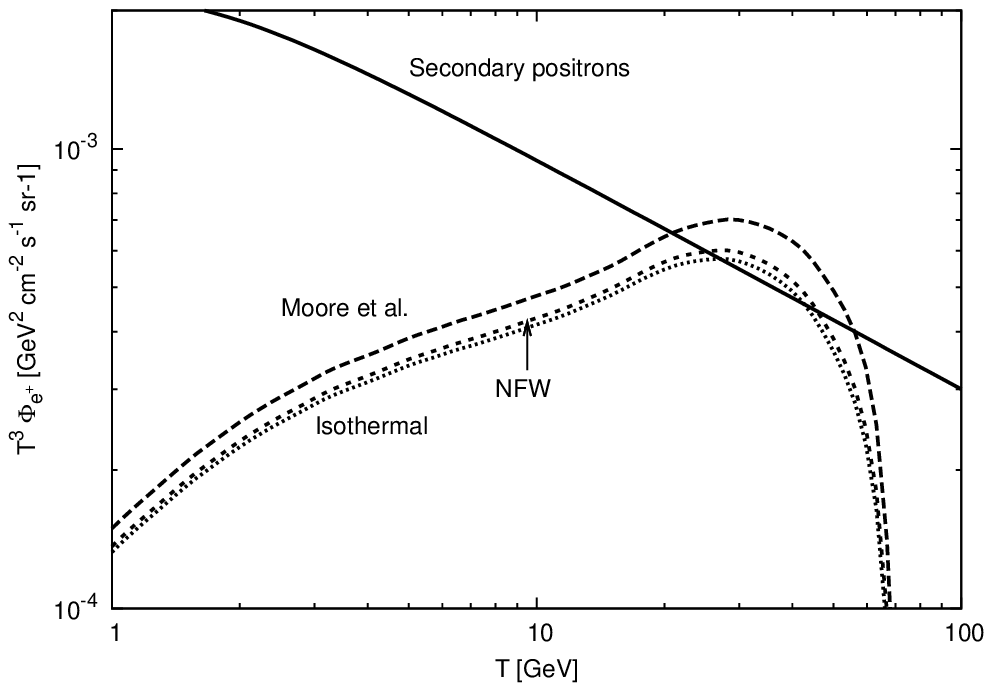,width=75mm} 
\psfig{figure=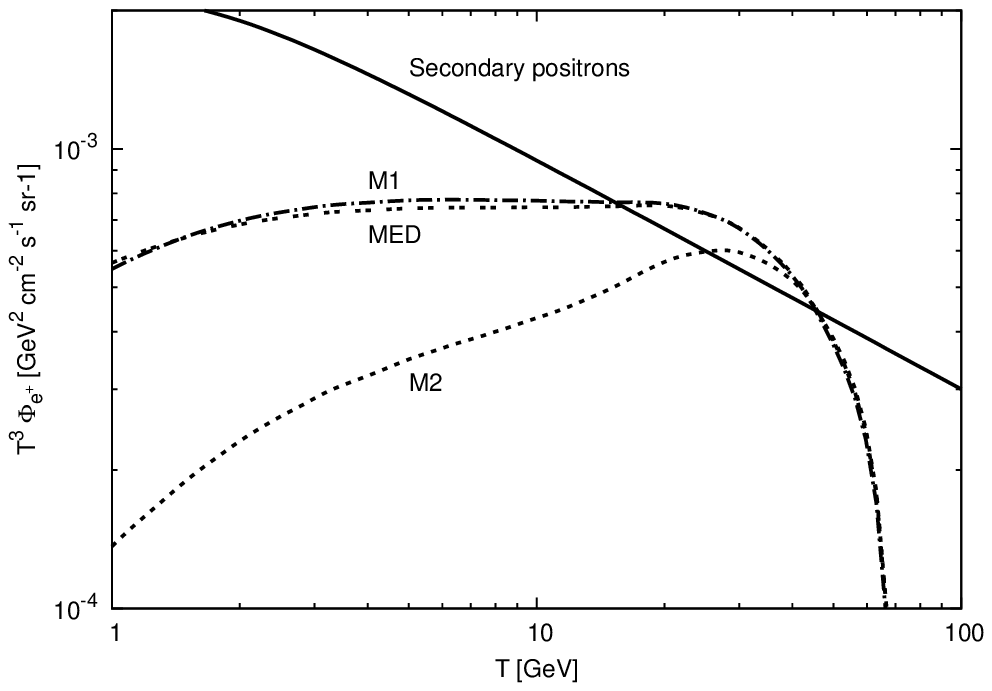,width=75mm}
\end{tabular}
\end{center}
\caption{\label{fig:pos-flux}\small 
Interstellar positron flux from the decay of gravitinos
with $m_{3/2}\simeq 150\,{\rm GeV}$ and 
$\tau_{3/2}\simeq 1.3\times 10^{26}\,{\rm s}$. In the left
plot we assume the M2 diffusion model (see Table \ref{tab:param-positron})
and we study the sensitivity of the positron flux to various
halo profiles. On the other hand, in the right plot
we assume a NFW halo profile and we study the sensitivity
of the positron flux to the diffusion model. We also show
for comparison the secondary positron flux from spallation
of cosmic rays on the Galactic disk.
}
\end{figure}

The interstellar positron flux from gravitino decay can 
be computed from Eqs.~(\ref{solution}) and (\ref{flux}), 
the result being:
\begin{equation}
\Phi_{e^+}^{\rm{prim}}(T) = \frac{c}{4 \pi m_{3/2} \tau_{3/2}} 
\int_0^{m_{3/2}}dT^\prime G_{e^+}(T,T^\prime) 
 \frac{dN_{e^+}(T^\prime)}{dT^\prime}\;.
\end{equation}
We show in Fig.~\ref{fig:pos-flux} the predicted interstellar 
positron flux from gravitino decay for different halo profiles
(left plot) and for different diffusion models (right plot). 
As expected, the dependence of the positron flux on the choice 
of the halo model is quite weak. On the other hand, the 
dependence on the diffusion model is important only at low
energies, where the signal lies well below the background.
At energies where the contribution to the
total positron flux from gravitino decay can be visible, $T\gsim 7$ GeV, 
the choice of the diffusion model only
changes the primary positron flux by a factor $2-3$.

Rather than measuring the positron flux, most experiments
measure the positron fraction, $\Phi_{e^+}/(\Phi_{e^-}+\Phi_{e^+})$,
since most sources of systematic 
error, such as detector acceptance or trigger efficiency,
cancel out when computing the ratio of particle fluxes.
Furthermore, the effects of solar modulation, which are important
in computing the positron flux at the top of the atmosphere 
below 10 GeV, also cancel out in the positron fraction when solar 
modulation is assumed to be charge-sign independent. 
In addition to the primary positron flux from gravitino decay there
exists a secondary positron flux originating from the collision of primary
protons and other nuclei on the interstellar medium, which
constitutes the background to our signal.
For the background fluxes of primary and secondary electrons, 
as well as secondary positrons, we use the parametrizations 
obtained in \cite{Baltz:1998xv} from detailed computer 
simulations of cosmic ray propagation~\cite{Moskalenko:1997gh}:
\begin{eqnarray}
\Phi_{e^-}^{\rm{prim}}(T) &=& \frac{0.16 \,T^{-1.1}}
{1 + 11 \,T^{0.9} + 3.2 \,T^{2.15}} 
~(\rm{GeV}^{-1} \rm{cm}^{-2} \rm{s}^{-1} \rm{sr}^{-1})\;,\\
\Phi_{e^-}^{\rm{sec}}(T) &=& \frac{0.70 \,T^{0.7}}
{1 + 110 \,T^{1.5} + 600 \,T^{2.9} + 580 \,T^{4.2}} 
~(\rm{GeV}^{-1} \rm{cm}^{-2} \rm{s}^{-1} \rm{sr}^{-1})\;,\\
\Phi_{e^+}^{\rm{sec}}(T) &=& \frac{4.5 \,T^{0.7}}
{1 + 650 \,T^{2.3} + 1500 \,T^{4.2}} 
~(\rm{GeV}^{-1} \rm{cm}^{-2} \rm{s}^{-1} \rm{sr}^{-1})\;,
\end{eqnarray}
where $T$ is expressed in units of GeV. Then, the positron fraction reads:
\begin{equation}
{\rm PF}(T) = \frac{\Phi_{e^+}^{\rm{prim}}(T) + \Phi_{e^+}^{\rm{sec}}(T)}
{\Phi_{e^+}^{\rm{prim}}(T) + \Phi_{e^+}^{\rm{sec}}(T) 
+ k \;  \Phi_{e^-}^{\rm{prim}}(T) + \Phi_{e^-}^{\rm{sec}}(T)},
\end{equation}
where following \cite{Baltz:1998xv,Baltz:2001ir} we have left 
the normalization of the primary electron flux
as a free parameter, $k$, to be fitted in order to match the 
observations of the positron fraction. When there is no
primary source of positrons, the positron fraction is best fitted for 
$k=0.88$~\cite{Baltz:2001ir}.

\begin{figure}[t]
\begin{center}
\begin{tabular}{c}
\psfig{figure=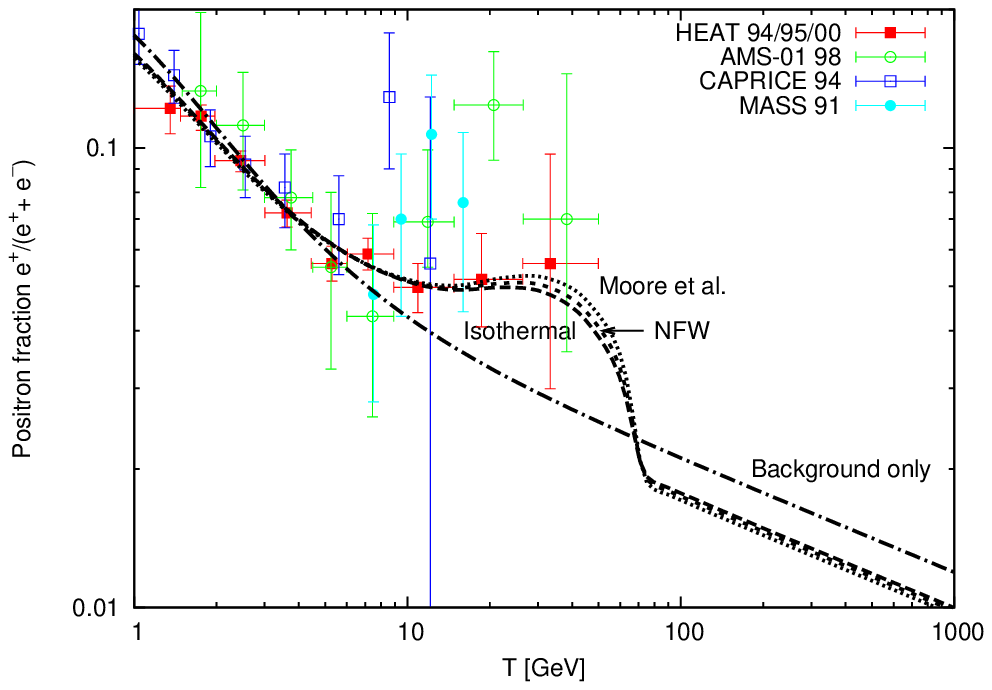,width=75mm} 
\psfig{figure=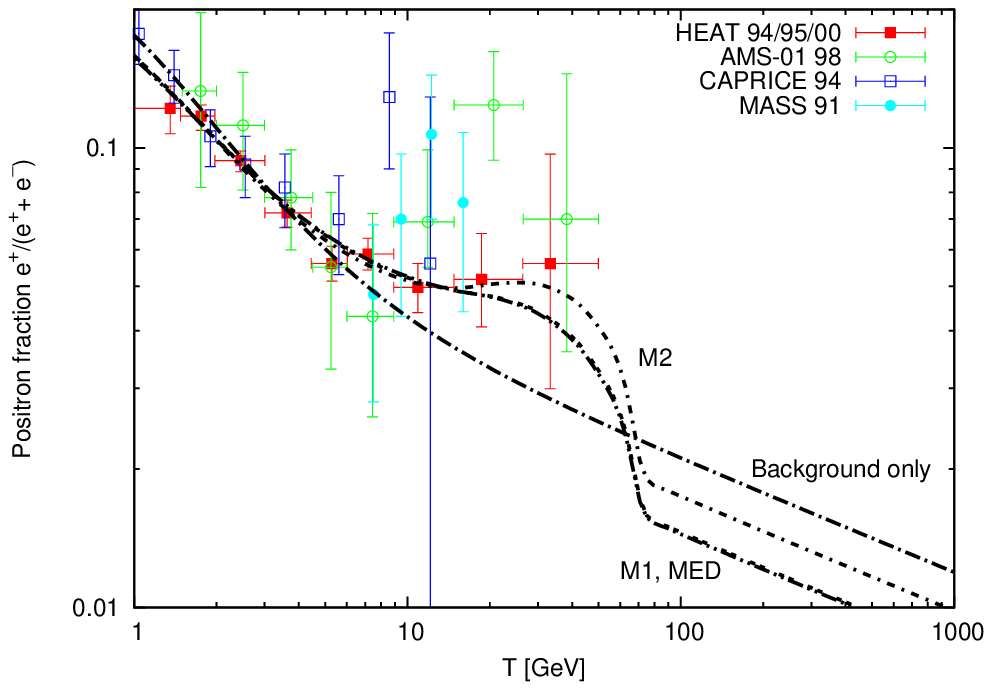,width=75mm}
\end{tabular}
\end{center}
\caption{\label{fig:pos-frac}\small 
Same as Fig.~\ref{fig:pos-flux}, but for the positron fraction.}
\end{figure}

We show in Fig.~\ref{fig:pos-frac} the positron fraction for 
different halo profiles (left plot) and for different diffusion 
models (right plot). In accordance with the results
for the primary positron flux, the dependence of the positron
fraction on the halo model is very weak. Furthermore,
the mild dependence of the primary positron flux on the choice
of the diffusion model becomes even milder when computing
the positron fraction. The reason for this is double: firstly,
the primary positron flux is never much larger than the secondary
positron flux, and secondly, the dependence on the choice
of diffusion model is partially absorbed by the normalization
of the primary electron flux that we have left as a free parameter.
Note also that the M2 model, which produces the minimal primary
positron flux, yields the most prominent bump in the positron 
fraction. This ``inversion'' is again a consequence of having left
the normalization of the primary electron flux as a free parameter. 
In order to reproduce the measured positron fraction at low energies, 
the normalization of the primary electron flux $k$ has to be
smaller in the M2 model than in the MED and M1 
(the precise values are $k=1.07,~1.28,~1.29$ 
for the M2, MED and M1 model respectively). Then, 
with the primary positron flux being comparable for all the diffusion
models at energies above $\sim 10$ GeV, the smaller value of $k$ for
the M2 model yields a larger positron flux in this energy range
than for the M1 and MED models.

In conclusion, we find that gravitino parameters which predict a departure
from a simple power law in the extragalactic gamma ray spectrum at energies
above 2 GeV (as observed by EGRET), {\it inevitably} predict a 
bump in the positron fraction at energies above 7 GeV (as observed
by HEAT). Furthermore, the presence of this feature is not very sensitive
to the many astrophysical uncertainties. 
This remarkable result holds not only for 
the scenario of gravitino dark matter with broken $R$-parity,
but also for any other scenario of decaying dark matter with lifetime
$\sim 10^{26}\,{\rm s}$ which decays
predominantly into $Z^0$ and/or $W^\pm$ gauge bosons with
momentum $\sim 50\,{\rm GeV}$.

\subsection{Antiproton Flux}

\begin{table}[t]
\begin{center}
\begin{tabular}{|c|cccc|}
\hline
Model & $\delta$ & $K_0\,({\rm kpc}^2/{\rm Myr})$ & $L\,({\rm kpc})$
& $V_c\,({\rm km}/{\rm s})$ \\
\hline 
MIN & 0.85 & 0.0016 & 1 & 13.5 \\
MED & 0.70 & 0.0112 & 4 & 12 \\
MAX & 0.46 & 0.0765 & 15 & 5 \\
\hline
\end{tabular}
\caption{\label{tab:param-antiproton} \small 
Astrophysical parameters compatible with the B/C ratio that 
yield the minimal (MIN), median (MED) and maximal (MAX) flux of antiprotons.}
\end{center}
\end{table}

The general transport equation, Eq.~(\ref{transport}), can be simplified 
by taking into account that energy losses are negligible for antiprotons. 
Therefore, the transport equation for the antiproton density, 
$f_{\bar p}(T,\vec{r},t)$, is simplified to:
\begin{equation}
0=\frac{\partial f_{\bar p}}{\partial t}=
\nabla \cdot (K(T,\vec{r})\nabla f_{\bar p})
-\nabla \cdot (\vec{V_c}(\vec{r})  f_{\bar p})
-2 h \delta(z) \Gamma_{\rm ann} f_{\bar p}+Q(T,\vec{r})\;,
\label{transport-antip}
\end{equation}
where the annihilation rate, $\Gamma_{\rm ann}$, is given by
\begin{equation}
\Gamma_{\rm ann}=(n_{\rm H}+4^{2/3} n_{\rm He})
\sigma^{\rm ann}_{\bar p p} v_{\bar p}\;.
\end{equation}
In this expression it has been assumed that the annihilation cross
section between an antiproton and a helium nucleus is
related to the annihilation cross section between an
antiproton and a proton by the simple geometrical factor $4^{2/3}$.
On the other hand, $n_{\rm H}\sim 1\;{\rm cm}^{-3}$ is the number
density of Hydrogen nuclei in the Milky Way disk,
$n_{\rm He}\sim 0.07 ~n_{\rm H}$ the number density
of Helium nuclei and $\sigma^{\rm ann}_{\bar p p}$ is
the annihilation cross section, which is parametrized 
by~\cite{Tan:1983de}:
\begin{eqnarray}
\sigma^{\rm ann}_{\bar p p}(T) = \left\{
\begin{array}{ll}
661\;(1+0.0115\;T^{-0.774}-0.948\;T^{0.0151})\; {\rm mbarn}\;,
 & T < 15.5\;{\rm GeV}~, \\
36 \;T^{-0.5}\; {\rm mbarn}\;, & T \geq 15.5\;{\rm GeV}\,, \\
\end{array} \right. 
\end{eqnarray}

Analogously to the positron case, the solution to the 
transport equation can be expressed as a convolution of the form 
Eq.~(\ref{solution}). The analytic expression for the 
Green's function reads~\cite{Donato:2001sr}:
\begin{equation}
G_{\bar p}(T,T^\prime)=\sum_{i=1}^{\infty}
{\rm exp}\left(-\frac{V_c L}{2 K(T)}\right)
\frac{y_i(T)}{A_i(T) {\rm sinh}(S_i(T) L/2)} 
J_0\left(\zeta_i \frac{r_{\odot}}{R}\right)\delta(T-T^\prime)\;,
\end{equation}
where
\begin{equation}
y_i(T)=\frac{4}{J^2_1(\zeta_i)R^2}
\int_0^R r^\prime \,dr^\prime\; J_0\left(\zeta_i \frac{r^\prime}{R}\right) 
\int_0^L dz^\prime {\rm exp}
\left(\frac{V_c (L-z^\prime)}{2 K(T)}\right)
{\rm sinh}\left(\frac{S_i(L-z^\prime)}{2}\right)
\rho(\vec{r}\,^\prime)\;,
\end{equation}
and
\begin{eqnarray}
A_i(T)&=&2 h \Gamma_{\rm ann}(T) + V_c+k S_i(T) 
{\rm coth} \frac{S_i(T) L}{2}\;,\\
S_i(T)&=&\sqrt{\frac{V_c^2}{K(T)^2}+\frac{4 \zeta_i^2}{R^2}}\;.
\end{eqnarray}
We find that the Green's function can be numerically 
approximated by the following interpolation function:
\begin{equation}
G_{\bar p}(T,T^\prime)\simeq 10^{14}\,
e^{x +y \ln T +z \ln^2T}
\delta(T^\prime-T)\,{\rm cm}^{-3}\,{\rm s}\;,
\label{interp-antip}
\end{equation}
which, again, is valid for any decaying dark matter particle. The coefficients
$x$, $y$ and $z$ for the NFW profile can be found in 
Table~\ref{tab:fit-antiproton}
for the various diffusion models in Table~\ref{tab:param-antiproton}.
In this case the approximation is accurate to a 5-10\%.
As in the case of the positrons, the dependence of the Green's
function on the halo model is fairly weak.

\begin{table}[t]
\begin{center}
\begin{tabular}{|c|ccc|}
 \hline
model & $x$ & $y$ & $z$ \\ 
\hline
MIN & $-0.0537$& 0.7052 & $-0.1840$\\
MED & 1.8002 & 0.4099 & $-0.1343$\\ 
MAX & 3.3602 & $-0.1438$ & $-0.0403$ \\
 \hline
\end{tabular}
\caption{\label{tab:fit-antiproton}\small 
Coefficients of the interpolating function Eq.~(\ref{interp-antip}) 
for the antiproton Green's function for the NFW halo profile.}
\end{center}
\end{table}

The interstellar antiproton flux is then given by
\begin{equation}
\Phi_{\bar p}^{\rm IS}(T)=\frac{v_{\bar p}(T)}{4\pi m_{3/2} \tau_{3/2}} 
\int_0^{m_{3/2}-m_p}dT^\prime G_{\bar p}(T,T^\prime) 
 \frac{dN_{\bar p}(T^\prime)}{dT^\prime}\;.
\end{equation}
However, this is not the antiproton flux measured by balloon
or satellite experiments, which is affected by solar modulation.
In the force field approximation~\cite{solar-modulation} 
the effect of solar modulation can be included
by applying the following simple formula that relates 
the antiproton flux at the top of the Earth's atmosphere and
the interstellar antiproton flux~\cite{perko}:
\begin{equation}
\Phi_{\bar p}^{\rm TOA}(T_{\rm TOA})=
\left(
\frac{2 m_p T_{\rm TOA}+T_{\rm TOA}^2}{2 m_p T_{\rm IS}+T_{\rm IS}^2}
\right)
\Phi_{\bar p}^{\rm IS}(T_{\rm IS}),
\end{equation}
where $T_{\rm IS}=T_{\rm TOA}+\phi_F$, with
$T_{\rm IS}$ and $T_{\rm TOA}$ being the antiproton kinetic energies 
at the heliospheric boundary and at the top of the Earth's atmosphere,
respectively, and $\phi_F$ being the solar modulation parameter,
which varies between 500 MV and 1.3 GV over the eleven-year solar
cycle. Since experiments are usually undertaken near
solar minimum activity, we will choose $\phi_F=500$ MV 
for our numerical analysis in order to compare our predicted flux with 
the collected data. 

\begin{figure}[t]
\begin{center}
\begin{tabular}{c}
\psfig{figure=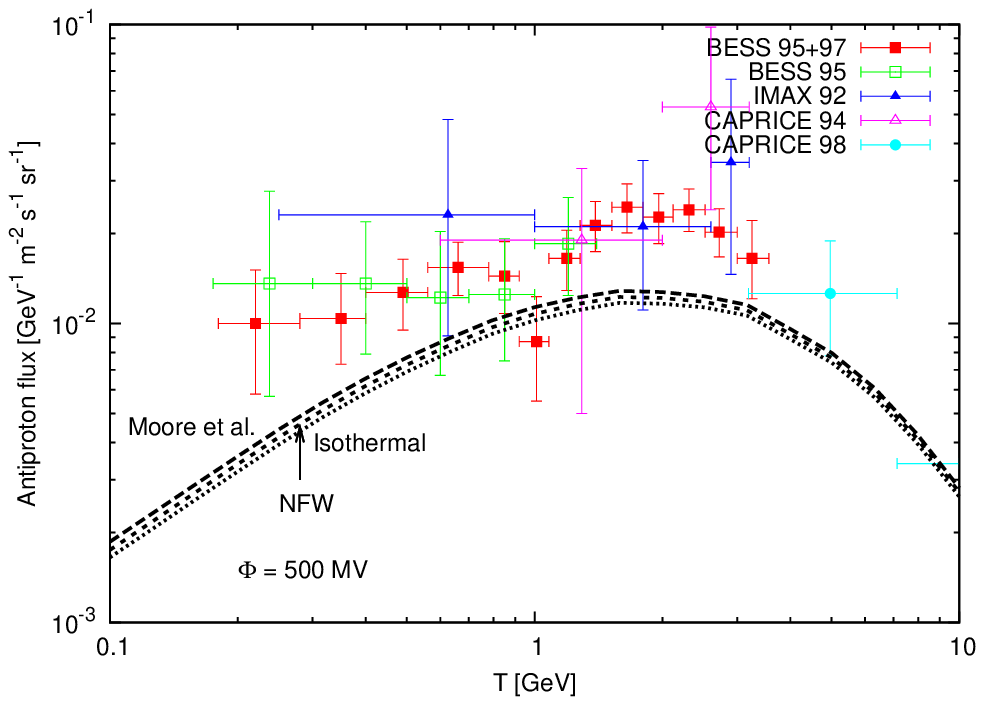,width=75mm} 
\psfig{figure=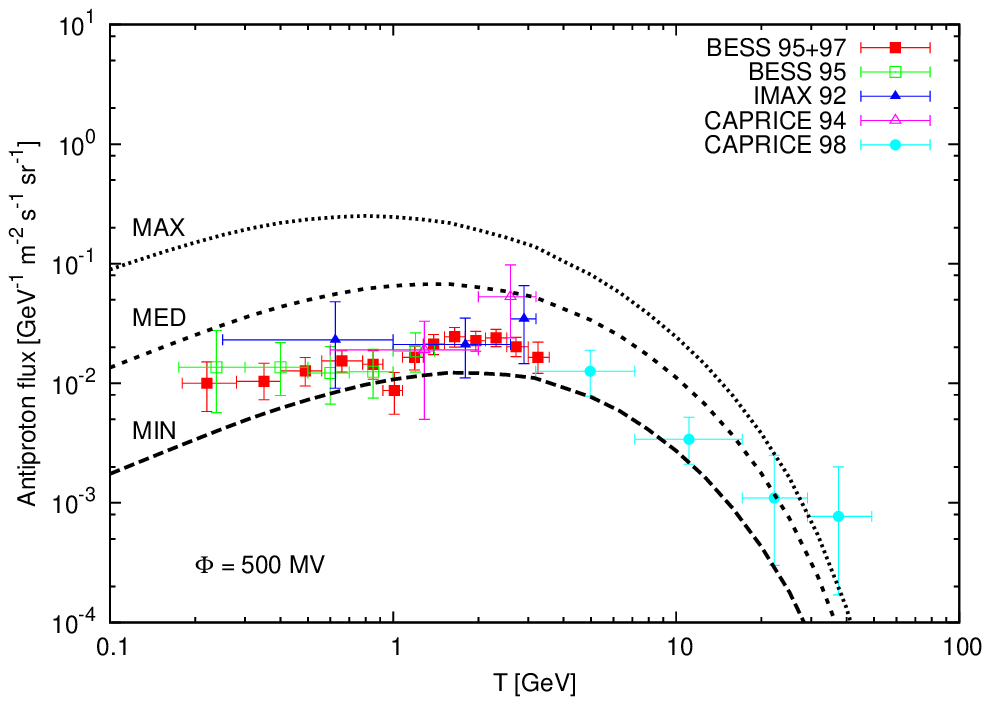,width=75mm}
\end{tabular}
\end{center}
\caption{\label{fig:antiproton}\small
Same as Fig.~\ref{fig:pos-flux}, but for the primary antiproton flux
at the top of the atmosphere. In the left plot the MIN diffusion 
model was assumed (see Table~\ref{tab:param-antiproton}).}
\end{figure}

We show in Fig.~\ref{fig:antiproton}, left plot, the predicted antiproton
flux for different halo models. As in the case of the positrons,
the sensitivity of the primary antiproton flux to the choice of halo model
is fairly mild.
We also show in the right plot the predicted antiproton
flux from gravitino decay for the diffusion models listed in 
Table~\ref{tab:param-antiproton}. From the plot,
the extreme sensitivity of the primary antiproton flux to the choice 
of the diffusion model is apparent: parameters that successfully reproduce the 
observed B/C ratio lead to antiproton fluxes that span over two
orders of magnitude. For a wide range of propagation parameters, the
total antiproton flux is well above the observations and thus
our scenario is most likely excluded, in spite of all the simplifying
assumptions in the diffusion model. However, the MIN model yields
a primary flux that is below the measured flux and thus might be
compatible with observations.

\begin{figure}[t]
\begin{center}
\begin{tabular}{c}
\psfig{figure=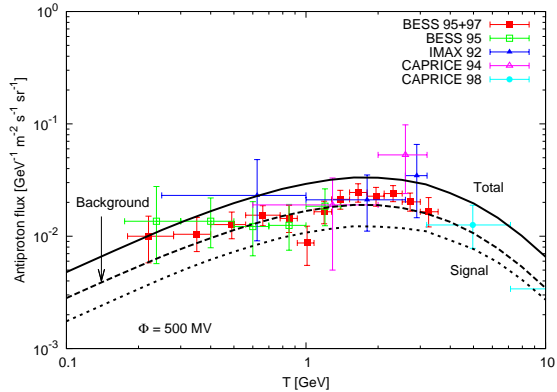,width=75mm}
\end{tabular}
\end{center}
\caption{\label{fig:antipMIN}\small
Contributions to the total antiproton flux in the MIN diffusion model.}
\end{figure}

We have analyzed more carefully the predictions for the MIN model
by computing the total antiproton flux. The result is shown in
Fig~\ref{fig:antipMIN}, where, for consistency, we have adopted as background
the secondary antiproton flux calculated in \cite{Donato:2001sr} 
for the same MIN model. Although the primary antiproton flux is
smaller than the measured one, the total antiproton flux is a factor
of two above the observations. Nevertheless, in view of all the 
uncertainties that enter in the calculation of the antiproton flux,
it might be premature to conclusively rule out the scenario of
decaying gravitino dark matter. Namely, in addition to the uncertainties
stemming from degeneracies in the diffusion parameters, there
are also uncertainties from the nuclear cross sections and, to a lesser
extent, uncertainties from the description of the interstellar
medium and solar modulation (for a discussion of the various uncertainties
see \cite{Donato:2001sr}). Furthermore, 
we used a simplified diffusion model that neglects the effects of
reacceleration, energy losses and tertiary contributions. 
Therefore, there could be certain choices of parameters
or more refined diffusion models where the total antiproton flux is
consistent with experiments\footnote{Some works have reported
a deficit in the predicted secondary antiprotons compared to the observations 
and argued that this deficit could be connected with a contribution of primary 
antiprotons~\cite{Moskalenko:2001ya}.}.

\section{Conclusions and Outlook}

In this paper we have calculated the positron and antiproton
fluxes from gravitino dark matter decay. The source term merely
depends on two parameters, the gravitino mass and the gravitino
lifetime, rendering a very predictive scenario from the particle
physics point of view. The main
uncertainties arise from the astrophysics, namely from our 
ignorance of the precise shape of the halo profile and especially 
from the degeneracies in the determination of the diffusion parameters.

By requiring a qualitatively good agreement of the predicted 
extragalactic gamma ray flux to the EGRET data, we have fixed  
$m_{3/2} = 150\,{\rm GeV}$ and 
$\tau_{3/2} = 1.3\times 10^{26}\,{\rm s}$. 
This choice of parameters completely fixes the source term,
and the only indeterminacy in the computation of the antimatter
fluxes stems from the unknown astrophysical parameters. 
Remarkably, with independence of the astrophysical uncertainties, we predict
a bump in the positron fraction at energies above 7 GeV, in 
agreement with the HEAT observations. On the other hand,
the predicted antiproton flux tends to be too large, 
although for certain choices of the propagation parameters the predicted
flux might also be in agreement with observations.

\begin{figure}[t]
\begin{center}
\begin{tabular}{c}
\psfig{figure=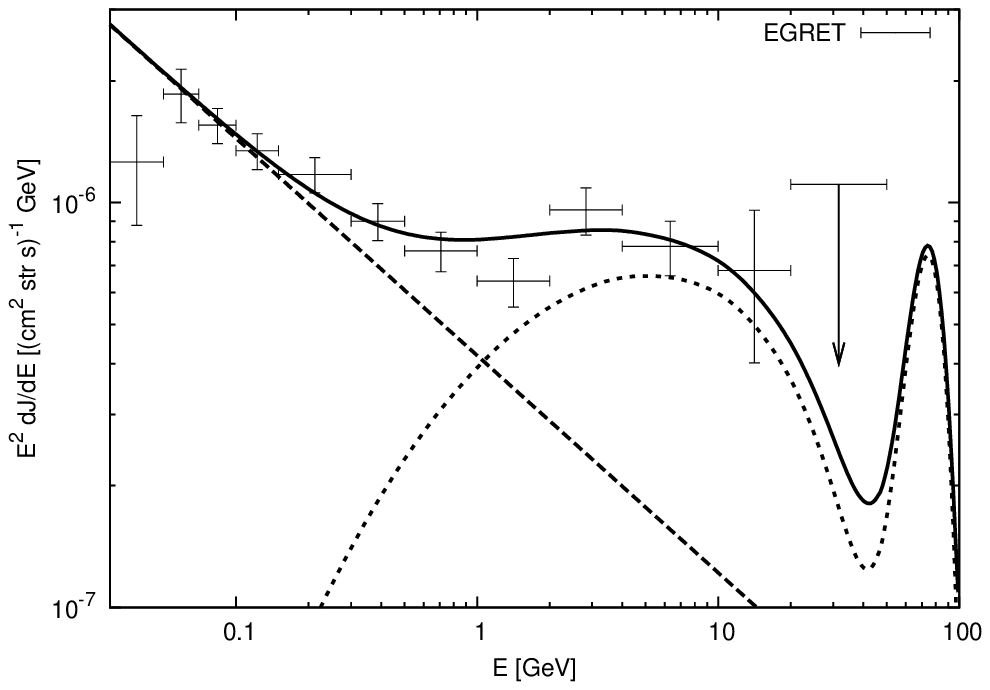,width=75mm} \\
\psfig{figure=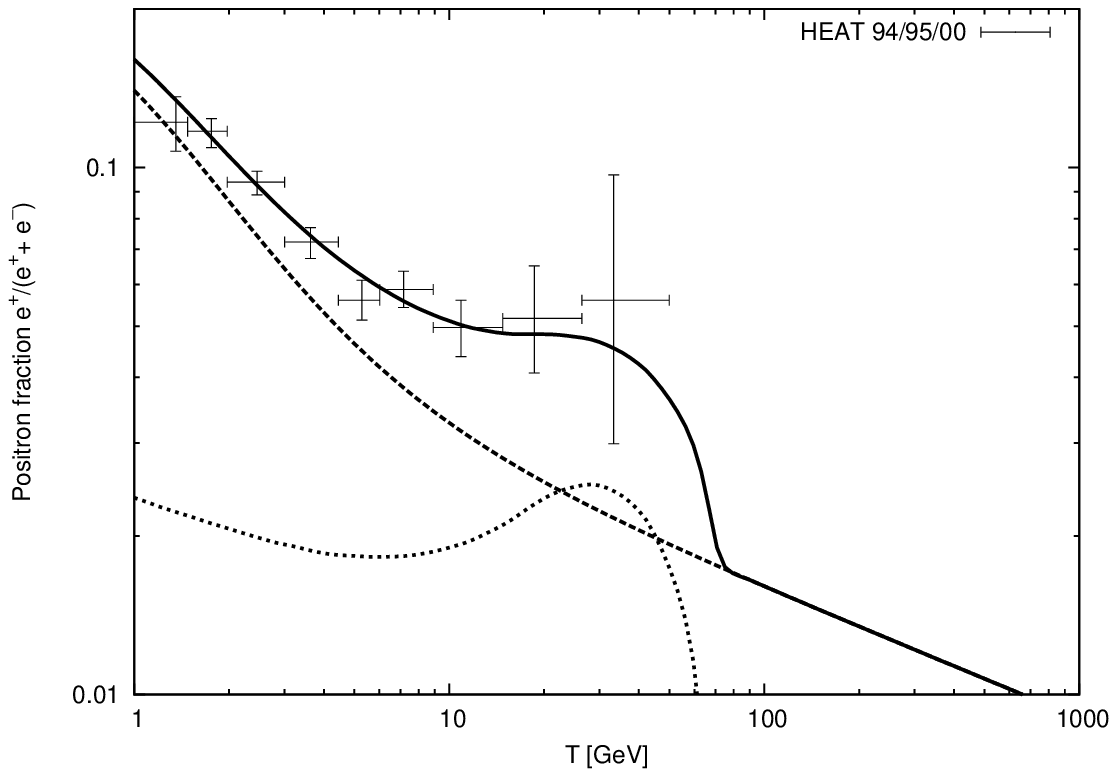,width=75mm}
\psfig{figure=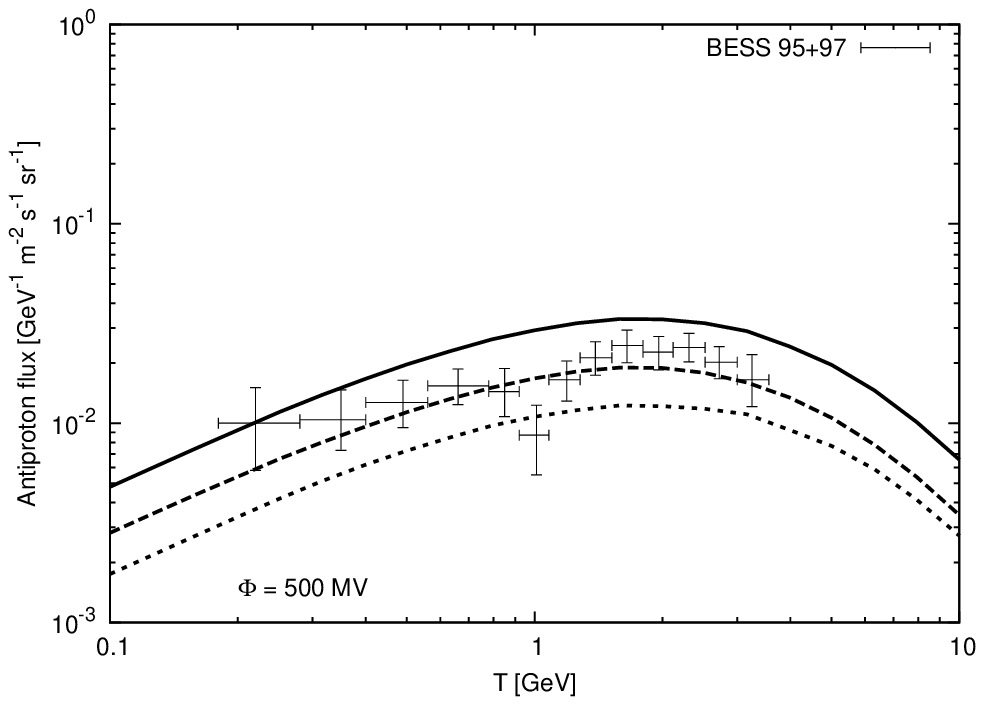,width=75mm}
\end{tabular}
\end{center}
\caption{\label{fig:summary}\small
Summary of the signatures of gravitino dark matter decay
in the extragalactic gamma ray flux (top), the positron fraction (bottom left)
and the antiproton flux (bottom right), compared to the EGRET, HEAT and
BESS data respectively. In these plots, we have adopted the MIN
diffusion model (see Table~\ref{tab:param-antiproton}),
$m_{3/2}\simeq 150\,{\rm GeV}$ and 
$\tau_{3/2}\simeq 1.3\times 10^{26}\,{\rm s}$. 
 }
\end{figure}

The main conclusion of this paper is summarized in the three 
plots in Fig.~\ref{fig:summary}. There, we show the predicted 
extragalactic gamma ray flux, positron fraction and antiproton flux 
compared to the EGRET, HEAT and BESS data respectively,
for $m_{3/2} = 150\,{\rm GeV}$, 
$\tau_{3/2} = 1.3\times 10^{26}\,{\rm s}$ and the
MIN diffusion model in Table~\ref{tab:param-antiproton}.
It is intriguing that for this diffusion model 
the scenario of gravitino dark matter with
broken $R$-parity can qualitatively explain the
anomalies observed in the extragalactic gamma ray flux {\it and} 
the positron fraction in a very natural way. 
It should also be stressed that this 
scenario was not devised to explain 
the anomalies in the cosmic ray fluxes, but to reconcile the 
clashing paradigms of supersymmetric dark matter, 
thermal leptogenesis and Big Bang nucleosynthesis.
At the same time, we find that the total antiproton
flux is slightly larger than the observed one. However, 
given all the uncertainties that enter in the calculation
of the antiproton flux, it might be premature to rule out
the present scenario on the basis of this small excess.

To conclude on the phenomenological viability of this scenario,
it would be worthwhile to elaborate
on the propagation of positrons and especially antiprotons
from gravitino decay by going beyond the simplified diffusion
model used in this paper~\cite{inprogress}. On the experimental side, 
the upcoming gamma ray experiment GLAST and the antimatter experiment
PAMELA will provide in the near future measurements of
the cosmic ray fluxes with unprecedented accuracy, thus providing
invaluable information about the scenario of decaying 
gravitino dark matter.

Finally, we would like to mention that these results are not
peculiar to the scenario of gravitino dark matter with broken
$R$-parity. The characteristic feature of our scenario is that 
the dark matter decays at late times into gauge bosons, eventually producing
photons, positrons and antiprotons in the fragmentation.
Therefore, similar signatures can be expected for other decaying dark
matter particles that couple to the $Z^0$ and/or $W^\pm$ gauge bosons.

\section*{Acknowledgements}

We are grateful to Wilfried Buchm\"uller, Marco Cirelli, 
Laura Covi, Michael Grefe and Piero Ullio for interesting
conversations.

\end{document}